\documentclass[twocolumn, prx, superscriptaddress, notitlepage]{revtex4-1}
\usepackage{graphicx}
\usepackage{physics}
\usepackage{float} 
\usepackage{subfigure} 
\usepackage{color}
\usepackage{amsmath}
\usepackage{amsfonts}
\usepackage[normalem]{ulem}
\usepackage{xspace}
\usepackage[dvipsnames]{xcolor}
\usepackage{dcolumn}
\usepackage[breaklinks,colorlinks,linkcolor=blue,citecolor=blue,urlcolor=blue]{hyperref}
\usepackage{lineno}

\newcommand{\be}{\begin{equation}}
\newcommand{\ee}{\end{equation}}
\newcommand{\bea}{\begin{eqnarray}}
\newcommand{\eea}{\end{eqnarray}}

\begin{document}
\title{Dipole condensates in synthetic rank-2 electric fields}

\author{Jiali Zhang}

 \affiliation{School of Physics, International Joint Laboratory on Quantum Sensing and Quantum Metrology, Center for Intelligence and Quantum Science, Huazhong University of Science and Technology, Wuhan 430074, China}

\author{Wenhui Xu}

 \affiliation{Department of Physics and Astronomy, Purdue University, West Lafayette, IN, 47907, USA}

%\author{Haizhou Lu}\email{luhaizhou@gmail.com} \affiliation{}

\author{Qi Zhou}
\email{zhou753@purdue.edu}
\affiliation{Department of Physics and Astronomy, Purdue University, West Lafayette, IN, 47907, USA}
\affiliation{Purdue Quantum Science and Engineering Institute, Purdue University, West Lafayette, IN 47907, USA}

\author{Shaoliang Zhang}
\email{shaoliang@hust.edu.cn}
\affiliation{School of Physics, International Joint Laboratory on Quantum Sensing and Quantum Metrology, Center for Intelligence and Quantum Science, Huazhong University of Science and Technology, Wuhan 430074, China}
\affiliation{Hubei Key Laboratory of Gravitation and Quantum Physics, Institute for Quantum Science and Engineering, Huazhong University of Science and Technology, Wuhan 430074, China}

\date{\today}

\begin{abstract}
Dipole condensates, formed from particle-hole pairs, represent a unique class of charge-neutral quantum fluids that evade conventional vector gauge fields, making their electrodynamic responses difficult to probe in natural materials.  Here, we propose a tunable platform using strongly interacting two-component ultracold atoms to realize dipole condensates and probe their coupling to rank-2 electric fields. By applying spin-dependent forces and treating spin as a synthetic dimension, we engineer a synthetic rank-2 electric field that induces measurable electrodynamic responses.  We identify the atomic analog of  perfect Coulomb drag: increasing intercomponent interactions leads to equal and opposite displacements of the centers of mass %dipole moments 
%in the two spin states.
of the two spin components.
Furthermore, a rank-2 electric field imprints a phase twist in the dipole condensate and generates a supercurrent of dipoles that obeys the dipolar Josephson relation---a smoking gun for dipole condensation. Our results establish a powerful platform for exploring dipolar superfluidity under tensor gauge fields.
\end{abstract}
\maketitle 

Whereas bosons tend to form a condensate at low temperatures, they may condense 
% in
by unusual means. Composite particles formed by multiple bosons—including particle-hole pairs—can themselves undergo condensation \cite{Law1998,SungKit2000,Mueller2006,evrard2021,Kokkelmans2022,Kokkelmans2022,Pu1998,Svistunov2003,Zhou2011,Zhang2021,Bigagli2024}. A primary example is the exciton condensates observed in a variety of solid or organic materials \cite{Spielman2000,eisenstein2004bose,Kasprzak2006,Balili2007,eisenstein2014,Kellogg2004,Tutuc2004,nandi2012,wang2019,li2017excitonic}. Another example is found  in the dipole Bose-Hubbard model, 
which supports
a dipolar condensate 
as the ground state 
in certain parameter regimes \cite{Rahul2022,Senthil2022,Senthil2023,Zechmann2023,Ye2023,Xu2024}. 
Additionally, certain non-equilibrium dynamics have been shown to dynamically generate dipole condensates \cite{Xu2024}.

Due to their charge-neutrality, dipoles are insensitive to ordinary vector gauge fields. For instance, a uniform electric field cannot induce a net dipole current, rendering dipole condensates effectively insulating \cite{Kohn1967,zeng2023}. To elicit electrodynamic responses, higher-rank tensor gauge fields are required \cite{Pretko2017,Chen2018,Barkeshli2018,Michael2017,Pretko2020,zhang2025}.  In the study of exciton condensates in bi-layer systems, Coulomb drag has been employed \cite{Kellogg2003,Narozhny2016,liu2017quantumhalldrag}. Applying an electric field to one layer induces a counterflow in the other due to the presence of strong interlayer Coulomb interactions. This phenomenon can be interpreted as a manifestation of a rank-2 tensor electric field \cite{Xu2024}.

Despite significant progress in understanding exciton condensates and Coulomb drag, fundamental challenges remain unresolved. One of the most pressing issues is the lack of %unambiguous evidence for exciton condensation, primarily due to the absence of 
direct detection of phase coherence in solids.  While transport measurements such as Coulomb drag suggest coherent behavior between electrons and holes, these signatures are often indirect and can be influenced by competing effects such as disorder. For instance, recent experiments reported the observation of perfect Coulomb drag but finite resistance remains \cite{qi2025perfect,nguyen2025perfect}.  Direct evidence for phase coherence of exciton condensates remains elusive. It is thus highly desirable to explore complementary platforms that offer greater control, cleaner environments, and more direct observables.

\begin{figure}[b] 
	\includegraphics[width=1\columnwidth]{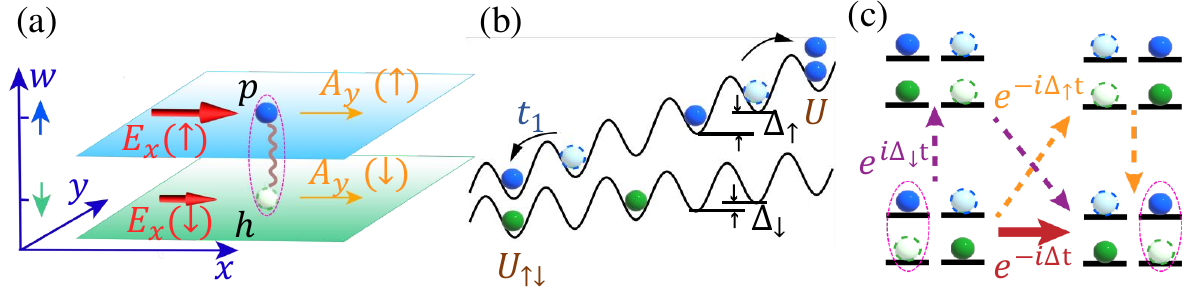}
	\caption{%Scheme of dipole condensates in tensor gauge fields. 
    (a) Spin-up (blue) and spin-down (green) atoms are regarded as two different layers in the synthetic dimension denoted by $w$. Spin-dependent effective electric field $E_x(\sigma)$ provides a dipole (ellipse) with a rank-2 electric field $E_{xw}$. %The dipole can be accelerated by rank-2 electric field which corresponds to the derivative of $E_x$ along $w$ direction. %(b) A rank-2 magnetic field which corresponds to the derivative of $B_{xy}$ along $w$ direction acts on a dipole.
    (b) An optical lattice with a spin-dependent linearly tilt %lattice with 
    $\Delta_{\uparrow}\neq \Delta_{\downarrow}$. %gives rises a finite $E_{xw}\sim $. 
    (c) The hopping of a dipole through second order processes $\sim e^{-i\Delta t}$ in the large $U$ limit, where $\Delta=\Delta_{\uparrow}-\Delta_{\downarrow}=E_{xw}$. 
    %that is equivalent to an electric field acting on a dipole
    }
    
	\label{fig:Scheme of tensor_field}
\end{figure}

In this manuscript, we propose that strongly interacting two-component atoms offer a versatile and controllable platform to investigate dipole condensates coupled to rank-2 electric fields.  The setup illustrated in Fig. \ref{fig:Scheme of tensor_field} consists of atoms with two internal states or two species of atoms, which are labeled by pseudospin $\sigma=\uparrow,\downarrow$.  With its high tunability and disorder-free environment, this setup enables unambiguous diagnosis of dipole condensation and allows for the controlled application of synthetic tensor gauge fields. To be more specific, in the presence of strong interactions, a single particle in neither spin could move. Instead, a particle from one component and a hole from the other form a dipole and a dipole condensate arises as the ground state. 
%A spin-dependent force is applied, as shown in Fig. \ref{fig:Scheme of tensor_field}(a). 
By interpreting the pseudospin as a synthetic dimension denoted by $w$, 
these two components 
are effectively separated from each other along $w$.
A spin-dependent force in the $x$-direction corresponds to an effective electric field $E_x({\sigma)}$, as shown in Fig. \ref{fig:Scheme of tensor_field}(a), and thus leads to a finite  %{\color{purple}inter-spin} 
gradient of $E_x$ in the $w$ direction. This gives rise to  
a discrete rank-2 electric field $E_{xw}= \partial_w E_x\sim E_x({\uparrow})-E_x({\downarrow})$.

Using this platform, we identify the atomic analog of the perfect Coulomb drag. Since it is more difficult to measure resistance than other quantities in atomic systems due to the absence of disorder, 
we demonstrate that tracking the displacement of the center of mass
% dipole moments 
of each spin component reveals Coulomb drag-like behavior. As the intercomponent interaction increases, the displacements of the centers of mass
% dipole moments
of spin-up and spin-down atoms become equal in magnitude and opposite in direction, mimicking perfect Coulomb drag.  Furthermore, we identify the smoking gun of the response of a dipole condensate to a rank-2 electric field. $E_{xw}$ imprints a phase twist on the dipole condensate, generating a supercurrent of dipoles. %while single-particle currents {\color{purple}are suppressed}.
% remain zero.
In particular, this supercurrent obeys the dipolar Josephson relation and is governed by the phase gradient of the dipole condensate, enabling a direct experimental measurement of phase coherence of dipole condensates \cite{Xu2024,Rontani2009}.

%We start by %creating a dipole condensate subject to a rank-2 tensor electric field. %in the absence of the tensor gauge fields. The %corresponding 
Since a uniform tensor electric field induces dynamics of dipoles only along one direction, it is sufficient to consider a 1D Hamiltonian, 
%It is sufficient to consider a one-dimensional system for simplification. The Hamiltonian of the one-dimensional bosons can be considered as 
\begin{equation}\label{full hamitonian}
\hat{H}_B = \hat{K}_B + \hat{U}_B+\hat{V}_B,
\end{equation}
where
\begin{eqnarray}\label{bhchainwithtwospins}
&\hat{K}_B &= -t_1\sum_{m,\sigma}(\hat{b}^\dag_{m,\sigma} \hat{b}_{m+1,\sigma} + h.c.), \\
&\hat{U}_B &= \frac{U}{2}\sum_{m,\sigma}\hat{n}_{m,\sigma}(\hat{n}_{m,\sigma}-1) + U_{\uparrow\downarrow}\sum_m\hat{n}_{m,\uparrow}\hat{n}_{m,\downarrow},\\
&\hat{V}_B& = \sum_{m,\sigma} m\Delta_\sigma \hat{b}^\dag_{m,\sigma}\hat{b}_{m,\sigma}.
\end{eqnarray} 
$\hat{b}^\dag_{m,\sigma}$($\hat{b}_{m,\sigma}$) is the bosonic creation (annihilation) operator %for bosons 
at site $m$ %with spin component 
for $\sigma = \uparrow,\downarrow$. $U$ and $U_{\uparrow\downarrow}$ correspond to the amplitude of intra-spin and inter-spin on-site interaction respectively, and $\hat{n}_{m,\sigma} = \hat{b}^\dag_{m,\sigma}\hat{b}_{m,\sigma}$. $\hat{V}_B$ denotes a spin-dependent linearly tilted potential as shown in Fig. \ref{fig:Scheme of tensor_field}(b).  %Such a model 
It is worth pointing out that $\hat{U}_B$ also describes bosons in the real space with long-range interactions. For instance, molecules or atoms with dipolar interactions could be placed in two separate layers \cite{du2024bilayerdipolar,tobias2022dipolar}.  $U$ then denotes the intra-layer onsite interaction and $U_{\uparrow\downarrow}$ denotes the inter-layer interaction. 

Applying a 
unitary %operator 
transformation, $\hat{\mathcal{U}}^\dagger\hat{H}_B\hat{\mathcal{U}}$, where $\hat{\mathcal{U}}=e^{i\hat{V}_Bt/\hbar}$, $\hat{V}_B$ is eliminated in the time-dependent Schr\"{o}dinger equation and the kinetic energy term $\hat{K}_B$ becomes time-dependent as $\hat{\tilde{K}}_B = -t_1\sum_{m,\sigma}(e^{-i\Delta_\sigma t/\hbar}\hat{b}^\dag_{m,\sigma} \hat{b}_{m+1,\sigma} + h.c.)$. This spin-dependent vector potential $A_\sigma=\Delta_\sigma t$ thus gives rise to a spin-dependent electric field $E_{\sigma}=\partial_t A_\sigma=\Delta_\sigma$,
%is constructed
%This leads to a finite gradient of the electric field in the
%synthetic $w$ dimension, i.e., 
and a finite $E_{xw}=E_{\uparrow}-E_{\downarrow}$. Whereas $\hat{K}_B+\hat{U}_B$ is the standard Hamiltonian of two-component atoms in optical lattices, a spin-dependent tilt has also been realized experimentally~\cite{Aidelsburger2013}.  %the tilted lattice can be achieved 
By adding a magnetic field with a homogeneous gradient along the $x$ direction, the Zeeman energy splitting between spin-up and spin-down atoms increases linearly with increasing $m$. A finite $\Delta/h$ is thus realized and its value could be tuned up to 10kHz, much larger than $t_1/h$, which is typically of the order of a few tens Hz or less. Our model in Eq.(\ref{full hamitonian}) is thus directly accessible in current experiments. 
%gradient can even be more larger than on-site interaction, so our discussion is relevant .

%What we are concerned about is in the 
%We are interested in the parameter regime, 
%regime of
%$0<t_1\ll V{\color{blue}<} U$ and %at half filling where 

We first consider the ground state of $\hat{H}_B$ when $\Delta_\sigma=0$. Previous studies have shown that the ground state is a dipole condensate, which was referred to as a counterflow superfluid in the parameter regime where $t_1\ll U,U_{\uparrow\downarrow}$ 
%at half filling 
\cite{Svistunov2003,Kuklov2004,Anzi2009}.
%{\color{purple} for instance, in the half-filled, repulsively interacting regime where the two-component $\nu$=1/2 per species, $U_{\uparrow\downarrow}>0$, dipole condensate realized for $t_1 / U \lesssim 0.16$ with the threshold $\left(U_{\uparrow\downarrow} / U\right)_{c} \simeq 32(t_1 / U)^{2}$ and below phase separation.}
A recent experiment has realized such a counterflow superfluid in optical lattices \cite{zheng2025}. Here, we numerically compute the correlation functions of the ground state using density matrix renormalization group (DMRG) method \cite{hauschild2018efficient}.
%(\ref{bhchainwithtwospins}). 
We choose $N=N_\uparrow + N_\downarrow = L$, where $N_\sigma$ 
is the particle number of spin $\sigma$, $N$ is total particle number, and $L$ is number of lattice sites. We observe that the one-body correlation function decays exponentially, $C_{mm'
,\sigma}=\langle \hat{b}^\dag_{m,\sigma} \hat{b}_{m',\sigma} \rangle\sim e^{-|m-m'|/\xi}$, where $\xi$ is the coherent length. Meanwhile, the reduced two-body density matrix decays as a power law,
\begin{equation}\label{dipolecorrelation}
C_{mm'}^D=\langle \hat{b}^\dag_{m,\uparrow} \hat{b}_{m,\downarrow} \hat{b}_{m',\uparrow} \hat{b}^\dag_{m',\downarrow}\rangle=\langle \hat{D}^\dag_m \hat{D}_{m'} \rangle\sim |m-m'|^\alpha,
\end{equation} 
where we define $\hat{D}^\dag_m = \hat{b}^\dag_{m,\uparrow}\hat{b}_{m,\downarrow}$ as the creation operator of a dipole at site $m$, and the exponent $\alpha$ depends on the interaction strength. For instance, when $U=2U_{\uparrow\downarrow} \gg t_1$, $\alpha$ approaches $-1/2$. These results confirm that the ground state here is indeed a dipole quasi-condensate in 1D.

\begin{figure}[htbp] 
	\includegraphics[width=1\columnwidth]{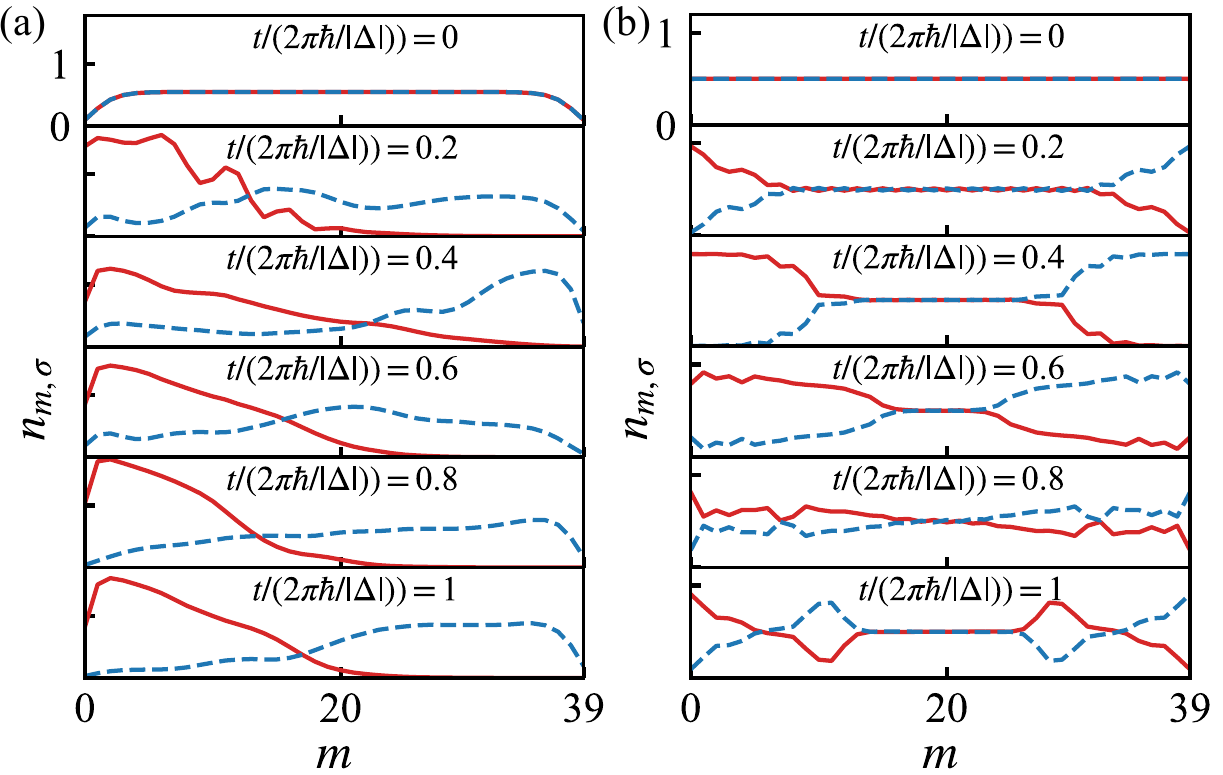}
	\caption{
    %(a)A spin-dependent linearly titled lattice gives rise to a finite $E_{xw}$ that is equivalent to an electric field acting on a dipole. (b) and (c) show the dipole correlation $C^D_{ij}$ and single-particle correlation $C_{ij,s}$ with distance $|i-j|$. The parameters are both chosen as $L=30$, $U=2V=80t_1$. To eliminate the impact of boundary, the central 20 lattice sites are shown. 
    Snapshots of the density profiles $n_{m,\sigma}(t)$ of spin-up (red solid line) and spin-down atoms (blue dashed line) in the presence of
    % The evolution of density distributions in Bloch oscillation by adding 
    a spin-dependent electric field with $\Delta_\uparrow = 0.05t_1$ and $\Delta_\downarrow = 0$. The size of the chain is $L=40$ with $N_\uparrow = N_\downarrow = L/2$. (a) and (b) illustrate weak and strong interactions, $U_{\uparrow\downarrow}=\frac{1}{2}U=0.25t_1$ and $10t_1$, respectively.
    % (a) $U=2U_{\uparrow\downarrow}=0.1t_1$ and (b) $U=2U_{\uparrow\downarrow}=20t_1$ {\color{purple}illustrate} the evolution of different interaction amplitudes, respectively. 
    %{\color{purple}Here $T=2\pi\hbar/|\Delta_{\uparrow}-\Delta_{\downarrow}|$ denotes the period of a dipole Bloch oscillation under the tensor electric field.}
    }
    
	\label{fig:coulomb drag}
\end{figure}

We then turn on a finite $\Delta_\uparrow$ while $\Delta_\downarrow$ remains zero. This mimics the Coulomb drag experiments in bilayer systems where a voltage is applied to only one layer. The subsequent dynamics is numerically simulated using the time-evolution block-decimation (TEBD) method \cite{hauschild2018efficient}. 
%{\color{purple}In our TEBD simulations we impose a local Hilbert space truncation at the maximum occupation number $N_\mathrm{max}=4$. The bond dimension $\chi_{max}$ is 400.} %We trace the subsequent dynamics of both spin components. 
Snapshots of the density profiles $n_{m,\sigma}(t)$ are shown in Fig. \ref{fig:coulomb drag}.  It is clear that spin-up atoms quickly moves to the left and their density profile changes significantly compared to the equilibrium result. %and 
%then bounce back and forth once the atoms reach the boundary. % signifying a Bloch oscillation of spin-up under a finite $\Delta_\uparrow$. 
In contrast, 
%spin-down atoms essentially remain stationary if 
%the movement of 
the density profile of spin-down atoms experiences only small changes %is not so obvious 
if $U_{\uparrow\downarrow}$ is small, %approaches $0$, %since the effective voltage applied to them %is {\color{blue} almost} 
%decreases to zero.
as shown in Fig. \ref{fig:coulomb drag}(a).
However, the results become drastically different once we increase ${U}_{\uparrow\downarrow}$. In spite of a vanishing $\Delta_\downarrow$, Fig. \ref{fig:coulomb drag}(b) shows that the density profile of the spin down atoms %first move toward right before colliding with the boundary.
also changes significantly compared to the equilibrium result. The increase (decrease) in the density of spin-up atoms is always accompanied by a symmetric decrease (increase) of spin-down atoms with respect to the equilibrium results. This is a direct consequence of the mutual interactions between spin-up and spin-down atoms that push spin-down atoms toward the opposite directions of spin-up atoms. Despite that atoms here interact with each other via onsite interaction, unlike excitonic systems where long-range Coulomb interactions exist, such correlated dynamics of spin-up and spin-down atoms is an atomic analog of the Coulomb drag.  

\begin{figure}[t] 
	\includegraphics[width=1\columnwidth]{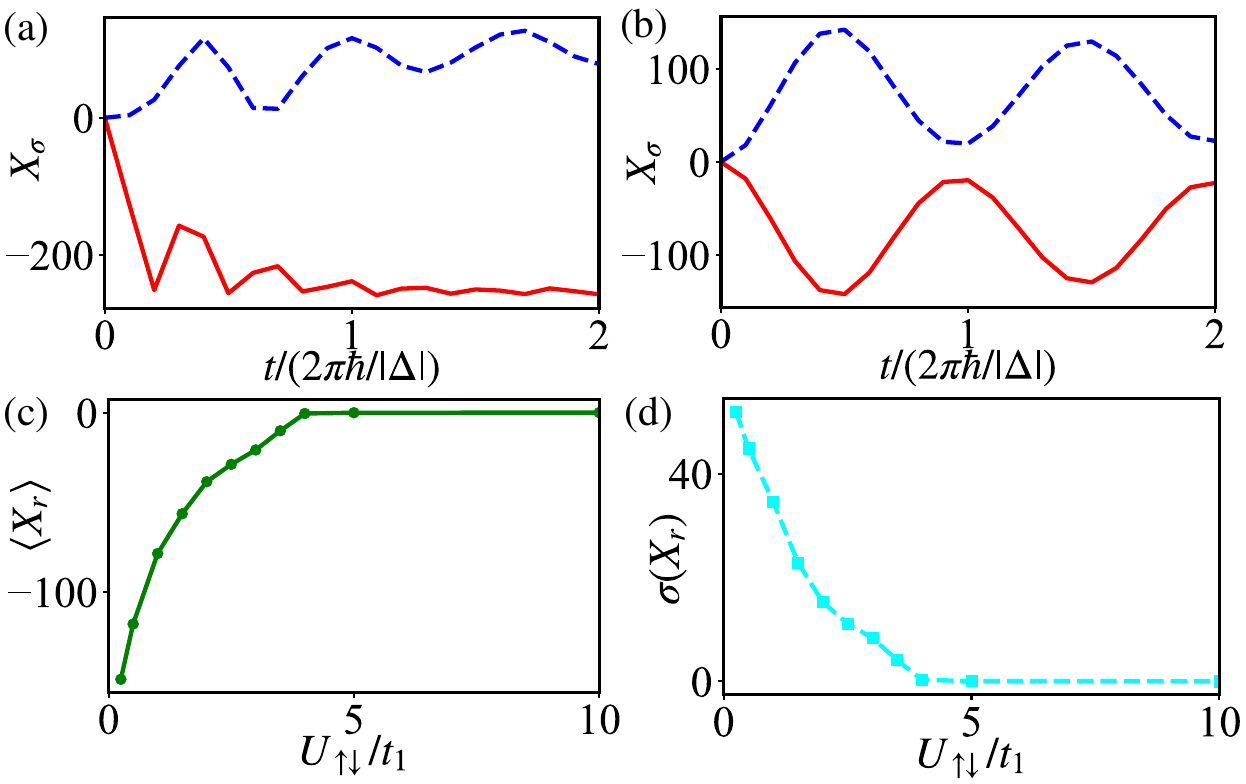}
	\caption{
    %(a) shows the evolution of amplitude of dipole correlation. Lines with different colors correspond to dipole correlation at different time, where $T=2\pi\hbar/|\Delta|$. (b) shows the evolution of phase of dipole correlation. Green solid line and yellow dot-dash line display the case with $\Delta=0.6t_1$ and $\Delta=0$ respectively. The parameters are chosen as $L=20$ and $U=2V=80t_1$. $\Delta=\Delta_\uparrow-\Delta_\downarrow = 0.6t_1$ is chosen for (a) and (b). (c) $n_k^D$ at different times. (d) The mean value of the momentum $\langle k\rangle$ as a function of time. 
    % The atomic analog of the Coulomb-like drag effect.
    (a,b) The displacement of the center of mass $X_{\sigma}$
    % dipole moments evolution
    of spin-up  (red solid line) and spin-down atoms (blue dashed line) 
    in the presence of a spin-dependent electric field with $\Delta_\uparrow = 0.05t_1$ and $\Delta_\downarrow = 0$ when  $U_{\uparrow\downarrow}=\frac{1}{2}U=0.25t_1$ (a) and  $U_{\uparrow\downarrow}=\frac{1}{2}U=10t_1$ (b). The size of the chain is $L=40$ with $N_\uparrow = N_\downarrow = L/2$.
    % components for different interaction amplitudes, where (a) $U=2U_{\uparrow\downarrow}=0.1t_1$ and (b) $U=2U_{\uparrow\downarrow}=20t_1$. The spin-dependent electric field is chosen as $\Delta_\uparrow=0.05t_1$ and $\Delta_\downarrow=0$ for both of these two cases. 
    (c,d) Time-averaged relative displacement $\langle X_r\rangle$ and its variance $\sigma({X_r})$
    % mean value of quadrupole moment for different interaction strengths
    as a function of the inter-spin interaction strength $U_{\uparrow\downarrow}$.}
    % The inset shows the fluctuation of the quadrupole moment. }
    
	\label{fig:dipole moment}
\end{figure}

To further quantitatively characterize the Coulomb-like drag in atomic systems, we evaluate the center of mass %dipole moment 
of each spin component, 
\begin{equation}
    X_\sigma=\sum_m m\langle \hat{n}_{m,\sigma}\rangle.
\end{equation}
%which represents the center of mass of a given spin component. 
%Discussions about the figure. 
In  Fig. \ref{fig:dipole moment}, we show the %derivation of dipole moments 
displacement of $X_\sigma$
from the initial value, $X_\sigma(t) - X_\sigma(0)$. 
% In Fig.\ref{fig:dipole moment} (a), the weak electric field $\Delta_\uparrow = 0.05t_1$ lead to the quick movement of spin-up dipole moment $P_\uparrow$ to a finite value.
%{\color{purple}In Fig.\ref{fig:dipole moment} (a), under 
%Whenthe weak electric field $\Delta_\uparrow = 0.05t_1$, the spin-up dipole moment 
In the presence of a finite $\Delta_{\uparrow}$, the center of mass of spin-up atoms
$X_\uparrow$ exhibits a rapid change, which is accompanied by  small oscillations due to a small but finite interaction with spin-down atoms $U_{\uparrow\downarrow}$, as shown in Fig. \ref{fig:dipole moment}(a).
%With no additional electric field, 
Meanwhile, due to the absence of an electric field acting on spin-down atoms, i.e., $\Delta_\downarrow=0$,
%the dipole moment of spin-down 
the change of $X_\downarrow$ %almost keeps constant. 
is much smaller, with a slight oscillation due to a finite but weak $U_{\uparrow\downarrow}$. 
%We should point out that, the oscillation of $P_\uparrow$ is not a signal of Bloch oscillation because the electric field is so weak that the density oscillation will be destroyed by the finite size effect. 
%But as in  Fig.\ref{fig:dipole moment} (b), with 
In sharp contrast, in the large $U_{\uparrow\downarrow}$ limit, the dynamics of spin-down atoms are highly correlated to spin-up atoms. As shown in Fig. \ref{fig:dipole moment}(b),
when $U_{\uparrow\downarrow}=\frac{1}{2}U=10t_1$, $X_\uparrow$ and $X_\downarrow$ oscillate with equal amplitudes and opposite signs. %simultaneously with opposite 
%{\color{purple}directions}. 
% value.
%In this case, the spin-up atom and on-site spin-down hole is bounded to move together, which corresponds to a perfect Coulomb-like drag. This is the signal of dipole Bloch oscillation because the amplitude of tensor electric field is not so small comparing with the tunneling amplitude of dipoles and the finite size effect has been suppressed. Because the dipole moment of {\color{purple}a spin-down hole is opposite in sign to that of a spin-up atom} as $-P_\downarrow$, one can define the quadrupole moment as

To analyze the correlated dynamics of the spin-up and spin-down atoms, we consider the relative displacement between the 
spin-up particles and spin-down holes.
% particles in spin-up atoms and the holes in spin-down atoms.
We note that for the same spin component, the center of mass of holes differs from that of particles by a minus sign. Thus, we define
\begin{equation}
    X_r=X_\uparrow - (-X_\downarrow) = X_\uparrow + X_\downarrow,
\end{equation}
which characterizes the relative displacement of the particles in the top layer from the holes in the bottom layer in the $x$ direction in Fig. \ref{fig:Scheme of tensor_field}(a). To quantitatively characterize how such displacement depends on $U_{\uparrow\downarrow}$ in the dynamics induced by $E_{xw}\sim \Delta_{\uparrow}-\Delta_{\downarrow}$, we define %Fig.\ref{fig:dipole moment} (c) shows the average {\color{purple}quadrupole moment} $Q$, 
the time-averaged $X_r$ and its variance: 
\begin{equation}
\begin{split}
       &\langle X_r\rangle =\frac{1}{t}\int^t_0 X_r(t') dt', \\
   &\sigma({X_r})=\left\{\frac{1}{t}\int^t_0[X_r(t')-\langle X_r\rangle]^2d t'\right\}^{1/2}.
   \end{split}
\end{equation} %{\color{purple}as a function of the interaction amplitudes} and the inset shows the 
%of $Q$ as 
With increasing $U_{\uparrow\downarrow}$, $\langle X_r\rangle$  approaches $0$, as shown in Fig. \ref{fig:dipole moment}(c). Furthermore, $\sigma({X_r})$ also vanishes in the large $U_{\uparrow\downarrow}$ limit, as shown in Fig. \ref{fig:dipole moment}(d). Since the integrand in the expression for $\sigma({X_r})$ is non-negative, a vanishing $\sigma({X_r})$ means that, once $\Delta_\uparrow$ drives oscillations of spin-up atoms, its center of mass is perfectly locked with the center of mass of holes in spin-down atoms at anytime, despite that spin-down atoms themselves are not subject to any finite $\Delta_{\downarrow}$. These results thus signify the atomic analog of perfect Coulomb drag in bilayer excitonic systems.  

%the fluctuation is also be $0$, which means that the quadrupole moment is conserved and the evolution corresponds to a perfect Coulomb-like drag. 

%In this case, 
%the large energy cost prevent the occupation of more than one bosons at each site. And the movement of 

Whereas the above results %of $P_\sigma$ 
allow us to explore  Coulomb-like drag in atomic systems, phase coherence and superfluidity of the dipoles cannot be directly extracted from such information. We thus explore the supercurrent induced by a phase twist applied to the dipole condensate. To this end, we apply a strong 
% {\color{green}$\Delta\equiv\Delta_\uparrow-\Delta_\downarrow$}
rank-2 tensor electric field $\Delta=\Delta_\uparrow-\Delta_\downarrow$
for a short duration, in which the density of the atoms does not change. To understand how this field
% method 
leads to a phase twist in the dipole condensate, we consider the effective model for dipoles in the strongly interacting regime.  
  
%{\color{green}As shown by Fig.(2a), the large penalty of interaction energy suppresses the single-particle kinetics and thus 
%single particle is also prevented because of the large energy detuning. Then the long-range off-diagonal order in the reduced single-article density matrix.} Meanwhile, %has been suppressed. But the 
As shown by Fig. \ref{fig:Scheme of tensor_field}(c), second-order processes %support the 
exchange atoms with different spins %components 
on the nearest neighbor sites\cite{Svistunov2003,Dai2017}. An effective Hamiltonian %can be derived with perturbation theory
is written as $\hat{H}_E = \hat{K}_E + \hat{U}_E$ where
\begin{eqnarray}\label{ham_dipolecondensate}
\begin{aligned}
\hat{K}_E =& -\sum_{m}(t_2e^{-i\Delta t/\hbar}\hat{b}^\dag_{m,\uparrow} \hat{b}_{m,\downarrow} \hat{b}_{m+1,\uparrow} \hat{b}^\dag_{m+1,\downarrow} + h.c. ), \\
\hat{U}_E =& \tilde{U}\sum_{m}(\hat{n}_{m,\uparrow}-\hat{n}_{m,\downarrow})(\hat{n}_{m+1,\uparrow}-\hat{n}_{m+1,\downarrow}),
\end{aligned}
\end{eqnarray}
$t_2=2t^2_1/U_{\uparrow\downarrow}$ is the amplitude of %ring
spin-exchange term, $\Delta = \Delta_\uparrow - \Delta_\downarrow$ is the rank-2 electric field, and $\tilde{U}=t^2_1(\frac{1}{U_{\uparrow\downarrow}}-\frac{2}{U})$ is the effective nearest-neighbor interaction strength \cite{Svistunov2003}. In the language of synthetic dimension, $t_2$ becomes a ring-exchange interaction in the $x-w$ plane. 
% We define $\hat{D}^\dag_m = \hat{b}^\dag_{m,\uparrow}\hat{b}_{m,\downarrow}$ as the creation operator of a dipole at site $m$,
The kinetic energy is rewritten as 
\begin{equation}\label{rank2electricfield}
\hat{K}_E = -\sum_{m}(t_2e^{-i\Delta t/\hbar}\hat{D}^\dagger_m \hat{D}_{m+1}+h.c.). %\hat{b}^\dag_{m,\uparrow} \hat{b}_{m,\downarrow} \hat{b}_{m+1,\uparrow} \hat{b}^\dag_{m+1,\downarrow}.
\end{equation}
In other words, a dipole sees a linearly growing vector potential or equivalently, a rank-2 electric field acting on a particle-hole pair.  

\begin{figure}[t] 
	\includegraphics[width=1\columnwidth]{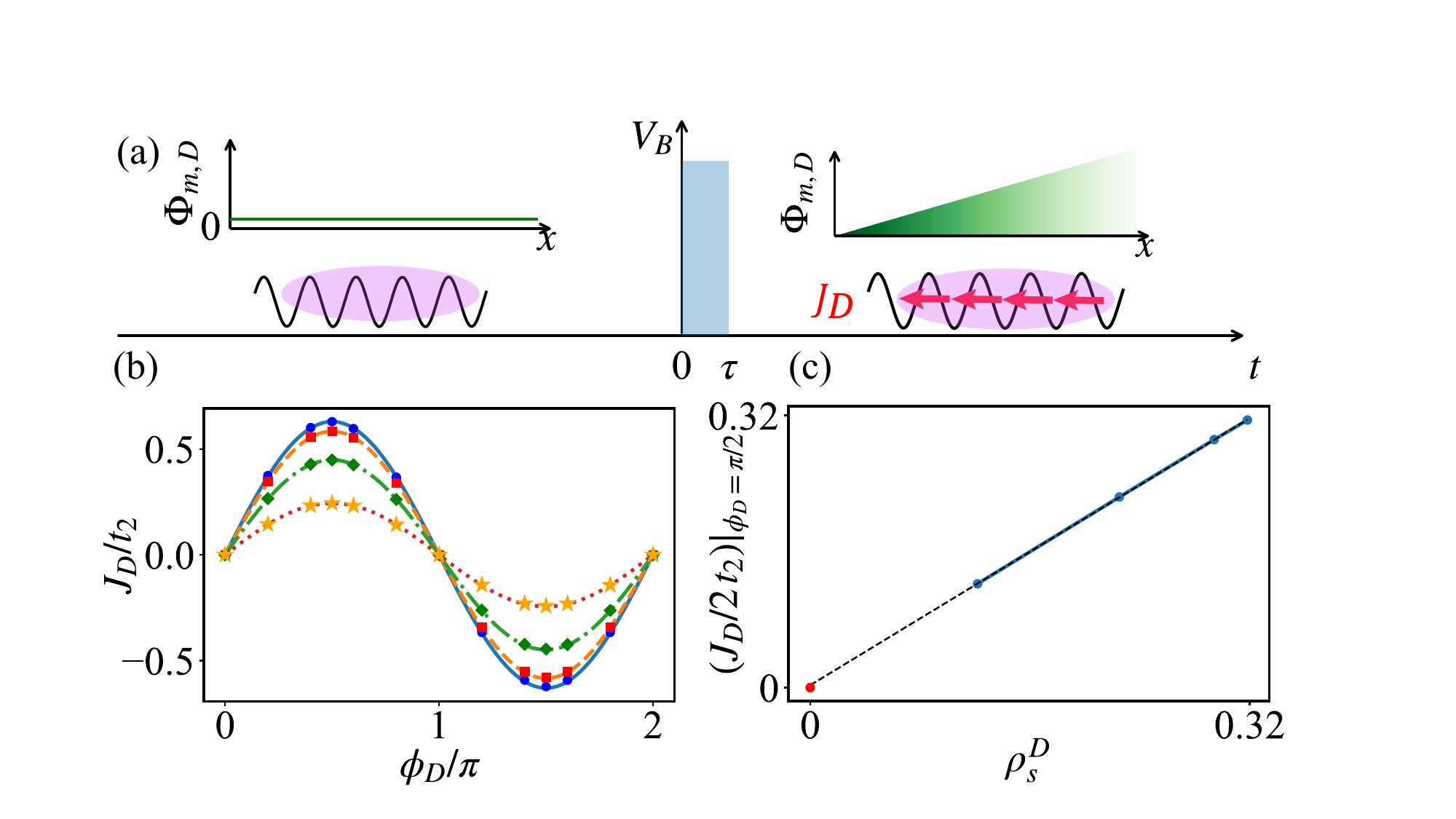}
	\caption{%(a) shows the evolution of amplitude of dipole correlation. Lines with different colors correspond to dipole correlation at different time, where $T=2\pi\hbar/|\Delta|$. (b) shows the evolution of phase of dipole correlation. Green solid line and yellow dot-dash line display the case with $\Delta=0.6t_1$ and $\Delta=0$ respectively. The parameters are chosen as $L=20$ and $U=2V=80t_1$. $\Delta=\Delta_\uparrow-\Delta_\downarrow = 0.6t_1$ is chosen for (a) and (b).  
	 (a) The scheme to generate dipolar supercurrents. A short pulse with amplitude  $\Delta=\Delta_\uparrow-\Delta_\downarrow$, applied between $t=0$ and $t=\tau$, generates 
    %additional 
    %spatial-dependent 
    a phase twist $\Phi_{m,D}$, which then induces  supercurrents of dipoles $J_D$. (b) The dependence of $J_D$ on the phase gradient $\phi_D$ at different dipole densities, $\rho = 0.125$ (orange stars), $\rho=0.25$ (green diamonds), $\rho=0.375$ (red squares) and $\rho=0.5$ (blue circles). Curves are fittings to sinusoidal functions.
    % Different color lines correspond to different 
    %particle
    % dipole }density: $\rho = 0.125$ (orange dashed dotted line), $\rho=0.25$ (green dotted line), $\rho=0.375$ (red dashed line) and $\rho=0.5$ (blue solid line). 
    (c) A comparison between %dipole superfluid density from 
    numerical results of $\rho^D_s$ and analytical results in Eq. (\ref{rhosrho}). }
    
	\label{fig:dipole supercurrent}
\end{figure}

Applying a strong rank-2 electric field to a dipole condensate for a short duration $\tau$ amounts to adding a phase to the ground state wavefunction $|G\rangle$, 
\begin{equation}
    |\Psi\rangle=e^{-i\sum_m m\Delta n_{m,D}\tau/\hbar}|G\rangle,
\end{equation}
where $n_{m,D}=\langle G|\hat{D}^\dagger_m\hat{D}_m|G\rangle$ is the density of dipoles at site $m$. The phase  $\Phi_{m,D} = m \Delta n_{m,D}\tau/\hbar$ %can be considered as 
is the consequence of the interplay between the dipole condensate and 
a linear potential $m\Delta$ generated by the rank-2 electric field, as shown in Fig. \ref{fig:dipole supercurrent}(a).  The phase difference of the dipole condensate between the $m$th and $(m+1)$th sites, i.e., the phase gradient, is written as 
\begin{equation}
    \phi_{m,D}=[(m+1)n_{m+1,D}-m n_{m,D}]\tau\Delta/\hbar.
\end{equation}
For a uniform density, $n_{m,D}=n_{D}$, the above expression can be simplified as 
\begin{equation}
    \phi_{D}= n_D\tau\Delta/\hbar.
\end{equation}
At the end of this short pulse, though the densities of both spin components remain essentially unchanged, a finite $\phi_{D}$ induces supercurrents of dipoles on the link between the $m$th and $(m+1)$th sites, 
\begin{equation}\label{dipolar Josephson}J_D  = 2t_2\rho^D_{s}\sin\phi_{D},
\end{equation}
where $\rho_{s}^D$ is the superfluid density of dipoles. This equation is referred to as the dipolar Josephson relation~\cite{Xu2024,Rontani2009}. 

To test the above analytical results, we numerically compute the supercurent between the $m$th and $(m+1)$th sites using the full model $\hat{H}_B$ in Eq. (\ref{full hamitonian}), 
\begin{equation}\label{dicurrent}
    J_D = 2t_2\mathrm{Im}\langle \hat{D}^\dag_m\hat{D}_{m+1} \rangle.
\end{equation}
% The results of $J_D$ immediately after the pulse is applied
The post-pulse values of $J_D$ are shown in Fig.  \ref{fig:dipole supercurrent}(b). Because of the small variation of $J_D$ across the whole system with nearly uniform density distribution, we have shown only the averaged $J_D$ of the whole system in this figure. We see clearly that $J_D$ is a sinusoidal function of $\phi_D$ and 
obeys the dipolar Josephson effect in Eq. (\ref{dipolar Josephson}). This is a hallmarking signature of a dipolar superfluid. %{(\color{red} If we define $J_D=\partial P_L/\partial t$, where $P_L=\sum_{j=1}^m \hat{n}_{j,\uparrow}-\hat{n}_{j,\downarrow}$, it has been verified by simulations of the full model in the large interaction regime that $J_D$ shows sinusoidal dependence on $\Delta_\uparrow$. But here $\partial_t P_L=-i[\hat{H}_B,P_L]=-2t \Im[\hat{b}_{m,\uparrow}^\dagger b_{m+1,\uparrow}-\hat{b}_{m,\downarrow}^\dagger b_{m+1,\downarrow}]$ is not equivalent to $\text{Im} \langle \hat{D}^\dagger_m \hat{D}_{m+1}\rangle$)}.
%{\color{blue}(The definition of $J_D$ should be $J_D = 2t_2 \mathrm{Im} \langle \hat{D}^\dagger_m \hat{D}_{m+1}\rangle$. Numerical calculations show that $j_{s,\uparrow}\approx j_{s,\downarrow} \approx j_D$)} %Meanwhile, we have verified that single-particle currents $J_{p,\sigma}=\text{Im} \langle \hat{b}^\dagger_{m,\sigma} \hat{b}_{m+1,\sigma}\rangle$ remain essentially zero. {\color{blue} 
%Eq.(\ref{rank2electricfield}) describes strongly correlated 1D bosons on a lattices when a particle-hole pair has been considered as a composite boson.

It is also worth mentioning that $\rho^D_s$ here has a simple
% simply
 analytical expression since dipoles can be treated as hard-core bosons in the large $U_{\uparrow\downarrow}$ limit. Furthermore, the nearest neighbor interaction in Eq. (\ref{ham_dipolecondensate}) can be made vanishingly small once $U_{\uparrow\downarrow}\rightarrow U/2$. %In this system, analytical calculation shows that the superfluid density should be 
%As such, the superfluid density of dipoles satisfy  
For hardcore bosons, the superfluid density satisfies 
$\rho_s
=\rho\sin(\pi\rho)/\pi$, where $\rho_s$ is the superfluid density and $\rho$ is the total density \cite{Rigol2005}. As such, $\rho_{s}^D$ here satisfies 
\begin{equation}
\rho_{s}^D
=\rho_D\sin(\pi\rho_D)/\pi,\label{rhosrho}    
\end{equation} where $\rho_D$ is the density of dipoles. 
%{\color{purple}dipoles.% formed by pairing a spin-up atom and a spin-down hole} 
To verify this relationship, we change $\rho_D=N_\uparrow/N$  by varying $N_\uparrow$ while keeping $N_\uparrow+N_\downarrow=L$ unchanged. We repeat the calculation of $J_D$ for various $\rho_D$ and extract $\rho_{s}^D$ using $J_D/(2t_2)$ at $\phi_D=\pi/2$ in Eq. (\ref{dipolar Josephson}). We then compare these results with $\rho_s^D$ obtained from the analytical relationship with $\rho_D$ in Eq. (\ref{rhosrho}). 
Fig. \ref{fig:dipole supercurrent} (c) shows that these two results agree well with each other, further demonstrating the validity of the dipolar Josephson relation.

%compares %that the superfluid density of dipoles,$\rho_D$ {\color{purple} %calculated 
%extracted from the numerical results of $J_D$ using Eq.(\ref{dipolar Josephson}) and that obtained from the
%exhibits excellent agreement 
%agrees well with 

%in our numerical calculations fit very well with these analytical results. }

We emphasize that in the strongly interacting limit, $J_D$ depends only on $\Delta=\Delta_{\uparrow}-\Delta_{\downarrow}$. We have verified that different choices of $\Delta_{\uparrow}$ and $\Delta_{\downarrow }$ at fixed $\Delta$ lead to the same $J_D$.
% We emphasize that $J_D$ depends on $\Delta$, not individual 
% $\Delta_\sigma$, in the strongly interacting limit, we have verified that different choices of $\Delta_{\uparrow}$ and $\Delta_{\downarrow }$ lead to the same $J_D$ provided that $\Delta=\Delta_{\uparrow}-\Delta_{\downarrow}$ is fixed. 
%{\color{blue} In current experiments, the tilted lattice can be achieved by adding a  magnetic field with homogeneous gradient along $x$ direction, the gradient can even be more larger than on-site interaction, so our discussion is relevant \cite{Aidelsburger2013}.}
When $\Delta_\uparrow=\Delta_\downarrow$, the rank-2 electric field vanishes and the system is subject to a uniform rank-1 electric field. We have found that $J_D$ vanishes, confirming that a dipole condensate is decoupled from a vector gauge field.

In ultracold-atom experiments, $J_D$  is directly observable. % \cite{Bloch2014chiralcurrents,Ruichao2024}. 
%{\color{green}Discussions about generalizing Yuan's experiment to measure $J_D$ and Bloch's and Alex Ruichao Ma's experiments on $J_{s,\sigma}$.}  
 %As in recent experiments in optical lattices or tweezer array, 
We define local spin operators $\hat{S}^z_{m} =(\hat{b}^\dag_{m,\uparrow}\hat{b}_{m,\uparrow} -\hat{b}^\dag_{m,\downarrow}\hat{b}_{m,\downarrow} )/2$, $\hat{S}^x_{m} = (\hat{b}^\dag_{m,\uparrow}\hat{b}_{m,\downarrow} + \hat{b}^\dag_{m,\downarrow}\hat{b}_{m,\uparrow} )/2$ and $\hat{S}^y_{m} = (\hat{b}^\dag_{m,\uparrow}\hat{b}_{m,\downarrow} - \hat{b}^\dag_{m,\downarrow}\hat{b}_{m,\uparrow} )/(2i)$. Since  $\hat{S}^{\pm}_m=\hat{S}^x_m \pm i\hat{S}^y_m$, $\hat{S}^+_{m}=\hat{b}^\dag_{m,\uparrow} \hat{b}_{m,\downarrow}$, $\hat{S}^-_{m}=\hat{b}^\dag_{m,\downarrow} \hat{b}_{m,\uparrow}$, %Since , 
the dipole correlation function $C_s=\langle \hat{D}^\dag_m \hat{D}_{m+1} \rangle=\langle \hat{S}^+_{m}\hat{S}^-_{m+1}\rangle$ is rewritten as 
% \begin{eqnarray}
\begin{equation}
\begin{aligned}
  C_s=\langle \hat{S}^x_{m}\hat{S}^x_{m+1} \rangle + \langle \hat{S}^y_{m}\hat{S}^y_{m+1}\rangle + i (\langle \hat{S}^y_{m}\hat{S}^x_{m+1} \rangle - \langle \hat{S}^x_{m}\hat{S}^y_{m+1} \rangle).\label{spincor}
\end{aligned}
\end{equation}
The dipole current in Eq. (\ref{dicurrent}) is then related to the spin-spin correlation functions, $J_D=2t_2 \text{Im}C_s$. %, where %rewritten as
%$\langle \hat{S}^z_{m,s} \hat{S}^z_{m,s}\rangle$ that amounts to the density-density correlations between two lattice sites can be directly obtained from in-situ density images.  
A recent experiment has measured the first two terms in Eq. (\ref{spincor})~\cite{zheng2025}. Since $\langle \hat{S}^z_m \hat{S}^z_{m+1}\rangle$ corresponds to the  density-density correlation $\langle \hat{n}_{m,\sigma}\hat{n}_{m+1,\sigma'}\rangle$, applying a global $\pi/2$ pulse, which rotates all spins about the $x$ ($y$) axis before the density images are taken, $\langle \hat{S}^y_{m} \hat{S}^y_{m+1}\rangle$ ($\langle \hat{S}^x_{m} \hat{S}^x_{m+1}\rangle$) is thus obtained. The last two terms in Eq. (\ref{spincor}) can also be measured using a straightforward generalization of this method. A site-selective $\pi/2$ pulse,  which rotates spins on even sites about the $x$-axis ($y$-axis) and other spins on odd sites about the $y$-axis ($x$-axis) before taking the density images, directly provides $\langle \hat{S}^y_{m}\hat{S}^x_{m+1} \rangle$ ($\langle \hat{S}^x_{m}\hat{S}^y_{m+1} \rangle$).

\begin{figure}[tbp] 
	\includegraphics[width=1\columnwidth]{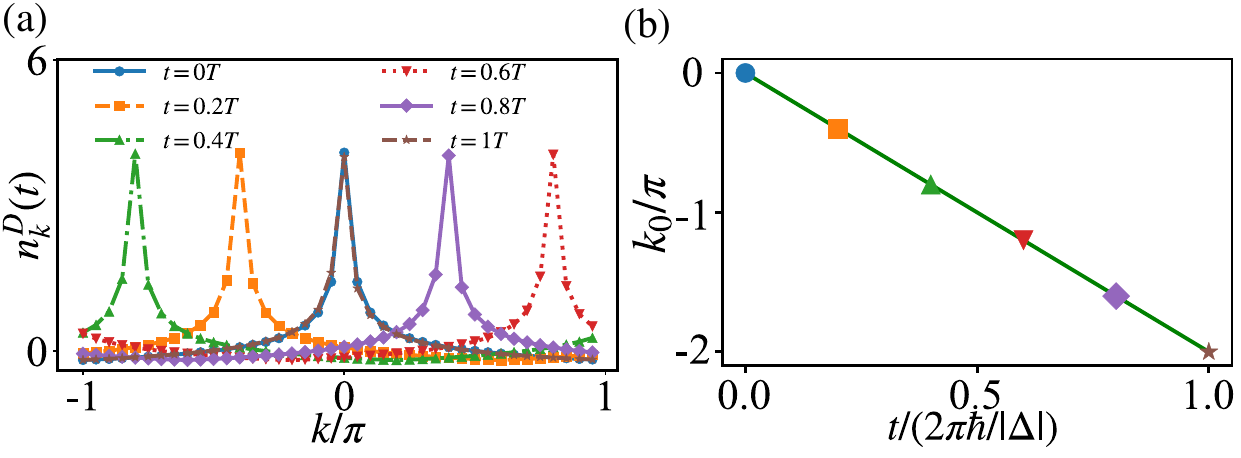}
	\caption{(a) %The scheme of a dipole condensate couple to rank-2 electric field. Blue and red dots correspond to spin-up and down bosons respectively. Linear tilted potential with $\Delta_\uparrow$ and $\Delta_\downarrow$ has been added on lattice with spin-up and down respectively. $t_2$ is the amplitude of dipole tunneling. (b) The density oscillation of dipole condensate in rank-2 electric field. (c) 
    The momentum distribution of dipoles at different times.
    % shows the evolution of amplitude of dipole correlation. 
    %Lines with different colors correspond to 
    % dipole correlation at different time,
    %different time steps, with the period 
     (b) The position of the maximum  $k_0$ %of the distribution shifts as 
     changes linearly with time.
    % shows the evolution of phase of dipole correlation. Green solid line and yellow dot-dash line display the case with $\Delta=0.6t_1$ and $\Delta=0$ respectively. 
    %The parameters are chosen as 
    $L=40$, $U_{\uparrow\downarrow}=\frac{1}{2}U=20t_1$, and $\Delta=\Delta_\uparrow-\Delta_\downarrow = 0.6t_1$ have been used. } 
	
	\label{fig:dipole_Bloch}
\end{figure}

In addition to the direct measurement of $\langle \hat{D}^\dag_m \hat{D}_{m+1} \rangle$, the dipole current $J_D$ can also be extracted from its continuity equation. If we use $J_D$ to denote the supercurrent of dipoles on the link  between the $j$th and the $(j+1)$th sites, $\langle\partial_t \hat{P}_L\rangle = -J_D$ is satisfied, where
$\hat{P}_L=\frac{1}{2}\sum_{m<j+1}\langle \hat{n}_{m\uparrow}-\hat{n}_{m\downarrow}\rangle$ is the total dipole moment of all sites on the left hand side of the link of interest ~\cite{Xu2024}. Measuring the time-dependent density distribution and the subsequent change rate in $\hat{P}_L$ then provide $J_D$.

An alternative means for studying the effect of rank-2 electric field is to consider the momentum distribution of dipoles. Similar to the momentum distribution of single-particles that is the Fourier transform of the single-body correlation function, we define the momentum distribution of dipoles, 
\begin{equation}
n_k^{D}(t)=\frac{1}{L}\sum_{m,m'}\langle \hat{D}^\dagger_m \hat{D}_{m'}\rangle e^{i k(m-m')}.%=\frac{1}{L}\sum_{i,j} C^D_{ij} e^{i k(i-j)}.
%|C^D_{r}| e^{ir\alpha  t}e^{-i r k}=\sum_r|C^D_{r}| e^{i r(\alpha  t-k)}. 
\end{equation}
As for the ground state, the momentum distribution is centered around $k=0$. %and $C^{D}\sim \sum_r|C^D_{r}|$, a characteristic feature of 1D quasi-condensate. 
After applying the strong pulse of $\Delta \gg t_2$,
the width of $n^D_k$ remains unchanged, and the position of the center, characterized by $k_0$ in the Brillouin zone, shifts when $t$ increases, as shown by Fig. \ref{fig:dipole_Bloch}(a). This is a signature of the Bloch oscillation of the dipole condensates induced by $E_{xw}$. The period of this dipolar Bloch oscillation is given by  $T=2\pi\hbar/|\Delta|$, as shown by Fig. \ref{fig:dipole_Bloch}. %In contrast, the momentum distribution of single-particles $n_{k\sigma}(t)=\frac{1}{L}\sum_{ij}\langle \hat{b}^\dagger_{i,\sigma}b_{j,\sigma}\rangle e^{ik(i-j)}$ remains essentially unchanged. 

%Meanwhile,  the phase of $C^{D}_{ij}$ changes linearly as a function of time $t$, i.e., $C^D_{ij}(t) \sim |C^D_{ij}| e^{i\alpha(i-j)t}$.

%The measurement of these spin correlations is accessible with the currently available experimental techniques.

In conclusion, we have shown that two-component atomic systems offer a highly tunable platform for studying the intriguing quantum phenomena induced by the coupling between dipole condensates and rank-2 electric fields. Pseudospin-dependent forces and strong interactions lead to the atomic analogy of perfect Coulomb drag and the dipolar Josephson effect, both of which are directly observable in current experiments. We hope that our results will stimulate more interest in studying the interplay of multipole condensates and higher-rank tensor gauge fields in synthetic quantum matter. 

%tensor gauge fields can induce abundant novel quantum phenomena, our scheme can also be generalized to higher-order multipole condensates and their coupling with corresponding higher-rank tensor gauge fields. 

{\bf Acknowledgements}:  QZ acknowledges supports from National Science Foundation (NSF) through Grant No. PHY-2110614. SZ is supported by National Natural Science Foundation of China (Grant No.12174138). The DMRG and TEBD simulations were performed by using the Tensor Network Python (TeNPy) package developed by J. Hauschild and F. Pollmann \cite{hauschild2018efficient} and were run on the HPC Platform of Huazhong University of Science and Technology.
\bibliographystyle{apstest}
\bibliography{dc.bib}

\end{document}